\documentclass[prl,aps,twocolumn,nopacs,floats,superscriptaddress]{revtex4-2}
\usepackage{float}
\usepackage{graphicx}
\usepackage{dcolumn}
\usepackage{bm}
\usepackage{amsmath}
\usepackage{amssymb}
\usepackage{color}
\usepackage{braket}

\usepackage{xcolor}

\definecolor{cardinal}{rgb}{0.6,0,0}
\definecolor{darkgreen}{rgb}{0,0.4,0}
\definecolor{golden}{rgb}{0.92, 0.7, 0}
\definecolor{midnight}{rgb}{0, 0, 0.5}
\definecolor{darkblue}{rgb}{0, 0, 0.7}

\def\he4{$^4$He}
\def\hel3{$^3$He}
\def\Am3{\AA$^{-3}$}
\def\beq{\begin{equation}}
\def\eeq{\end{equation}}

\newcommand{\be}{\begin{equation}}
\newcommand{\ee}{\end{equation}}
\newcommand{\bea}{\begin{eqnarray}}
\newcommand{\eea}{\end{eqnarray}}
\newcommand{\bse}{\begin{subequations}}
\newcommand{\ese}{\end{subequations}}

\begin{document}
\title{Emergent BEC mechanism in flat-band superconductors}

\author{V. Berger}
\affiliation{Department of Physics, University of Massachusetts, Amherst, MA 01003, USA}
\author{I.S. Tupitsyn}
\email{itupitsyn@physics.umass.edu}
\affiliation{Department of Physics, University of Massachusetts, Amherst, MA 01003, USA}
\author{B. Currie}
\affiliation{Physics Department, King's College London, Strand, London WC2R 2LS, UK}
\author{B.V. Svistunov}
\affiliation{Department of Physics, University of Massachusetts, Amherst, MA 01003, USA}
\affiliation{Physics Department, King's College London, Strand, London WC2R 2LS, UK}
\author{E. Kozik}
\affiliation{Physics Department, King's College London, Strand, London WC2R 2LS, UK}
\author{N.V. Prokof'ev}
\affiliation{Department of Physics, University of Massachusetts, Amherst, MA 01003, USA}
\affiliation{Physics Department, King's College London, Strand, London WC2R 2LS, UK}

\begin{abstract}
The formation of bound bosonic pairs of fermions, followed by their Bose-Einstein (quasi)  condensation (BEC), is a foundational mechanism of superconductivity. At low filling, flat-band superconductivity is well captured by this mechanism provided the flat band is separated from the occupied lower band by an energy gap. However, particularly high $T_c$ values are anticipated when the non-interacting flat and lower bands touch---as in the prototypical attractive Lieb-lattice model studied here—--invalidating the conventional picture: while interactions might protect the bound state by opening a gap, no small parameter guarantees the separation of the bound-state energy from this gap or occupied-band excitations, leaving the pair's fate uncertain. Based on a controlled-precision numerical protocol---which we demonstrate to be essential in this fundamentally non-perturbative problem---we show that the BEC mechanism, underpinned by an interaction-induced gap, is generically robust and remarkably efficient: fermions doped into the flat band form bound pairs within this gap with an anomalously light effective mass, enabling an exceptionally high $T_c$.
\end{abstract}

\maketitle


\textit{Introduction.}
In systems of non-interacting lattice fermions, destructive interference between wavefunction amplitudes on different lattice sites can obstruct particle propagation and result in flat (dispersionless) Bloch energy bands. Since the effective mass of a particle in a flat band (FB) is infinite, one might think that in this situation interactions will force FB fermions into an insulating charge density wave state. Rather counter-intuitively, FB can also host a superconducting state at sufficiently low temperature $T<T_c$. Moreover, the transition temperature, $T_c$, may be \textit{enhanced} by the presence of FB \cite{peotta_superfluidity_2015, julku_geometric_2016} because at weak negative-$U$ Hubbard coupling the theory predicts linear dependence of $T_c$ on $|U|$ \cite{kopnin_high-temperature_2011}. Numerous early studies of FB superconductivity focused on models with isolated FBs \cite{peotta_superfluidity_2015, julku_geometric_2016, liang2017band, tovmasyan2016effective,Berg2020, Berg2023}. Subsequent work investigated gapless models \cite{huhtinen_revisiting_2022, iskin_origin_2019}
and found qualitatively similar results, while the most recent results 
\cite{penttila_flat-band_2025, tupitsyn_numerically_2026} identified gapless (at the level of a non-interacting Hamiltonian) systems as having the strongest superconducting response.

When the band structure features a large gap, $\Delta \gg |U|$, between the FB and the lower occupied bands (if any), one can begin by solving the two-body problem in the empty-bands subspace to determine whether two opposite-spin fermions form a bound state and what its properties are. It has been shown that such bound states have a finite effective mass $m^*$ \cite{torma_quantum_2018, iskin_two-body_2021}, which is tied to the quantum metric \cite{liu2025quantum} and the strong momentum dependence of the interaction Hamiltonian. This explains why the bound-state energy depends on the total pair momentum, allowing pairs to remain mobile despite the infinite mass of their single-particle constituents. Bound states are more than just a conceptual demonstration---when their density, $n_b$, is low enough (pair size is smaller than the distance between pairs) they can be treated as point bosons and used to accurately estimate $T_c$ from the relation
\cite{zhang_superconducting_2023}
\bea
T_c\approx 1.3 \, (n_b/m^*) = 0.65 (n_f/m^*) \, ,
\label{Tc}
\eea
where $n_f=2n_b$ is the density of fermions doped into FB. As discussed in Ref.~\cite{zhang_superconducting_2023}, this formula works in a very broad range of bosonic densities irrespective of the microscopic model details (including the interaction range).
This is how a complex problem of superconductivity in FB systems is radically simplified and reduced to computing properties of the bound state.

However, without a gap $\Delta$ separating the flat band from the lower occupied band in the non-interacting Hamiltonian, two-body bound states in the flat band with binding energy $E_b < 0$ are ill-defined: their energy per particle would overlap with the lower band, triggering a many-body reconstruction of the entire spectrum. Even with $\Delta > 0$, the stability of the insulating ground state---above which bound pairs form a distinct branch of elementary excitations---requires the bound-state energy per particle to lie above the top of the occupied band, corresponding to a positive gap $\delta = E_b + 2\Delta > 0$ in the two-particle channel.
It is then natural to question whether a well-defined charge-2$e$ quasiparticle even \textit{exists} in a system where the bare dispersion is such that FB has no gap from below. If not, the resulting superconducting ground state would be distinctly different from the one formed by the BEC of bosonic pairs, invalidating the physical picture leading to Eq.~(\ref{Tc}). In the absence of the small parameter $|U|/\Delta \ll 1$, predicting the precise physical scenario becomes challenging.

In this Letter, we present analytic arguments and numeric evidence for the existence of a stable, dispersive charge-2$e$ quasiparticle in the standard Lieb lattice that has no gap between the flat and occupied lower band in the non-interacting system. Using diagrammatic schemes of increased complexity, from two-body solutions in restricted Hilbert space to beyond mean-field diagrammatic schemes, and, ultimately, a numerically exact solution by diagrammatic Monte Carlo (DiagMC), we show that the attractive on-site interaction opens up a band gap between the filled lower band and FB. When singlet fermion pairs are doped into this \textit{interaction-induced band insulator} (IIBI)  they form bound states with energies (per particle) within the band gap. By computing the effective mass of the bound pairs, we arrive at an accurate parametrization of the superconducting transition temperature at low density $n_f$.

\begin{figure}[t]
\centering
\raisebox{0.22\height}{
\includegraphics[width=0.42\linewidth]{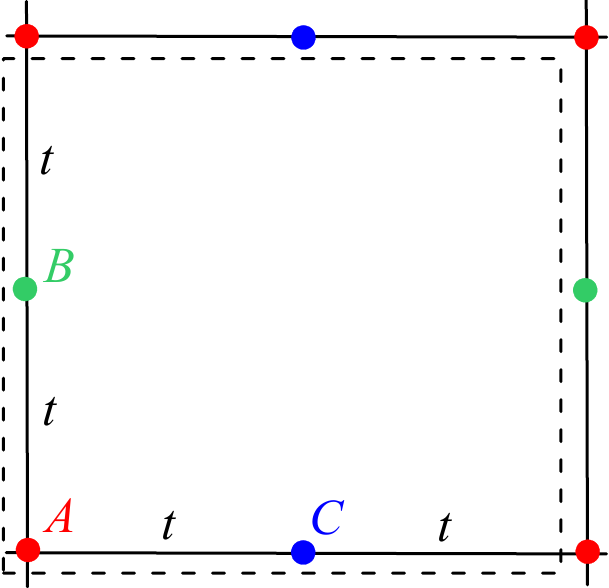}}
\includegraphics[width=0.55\linewidth]{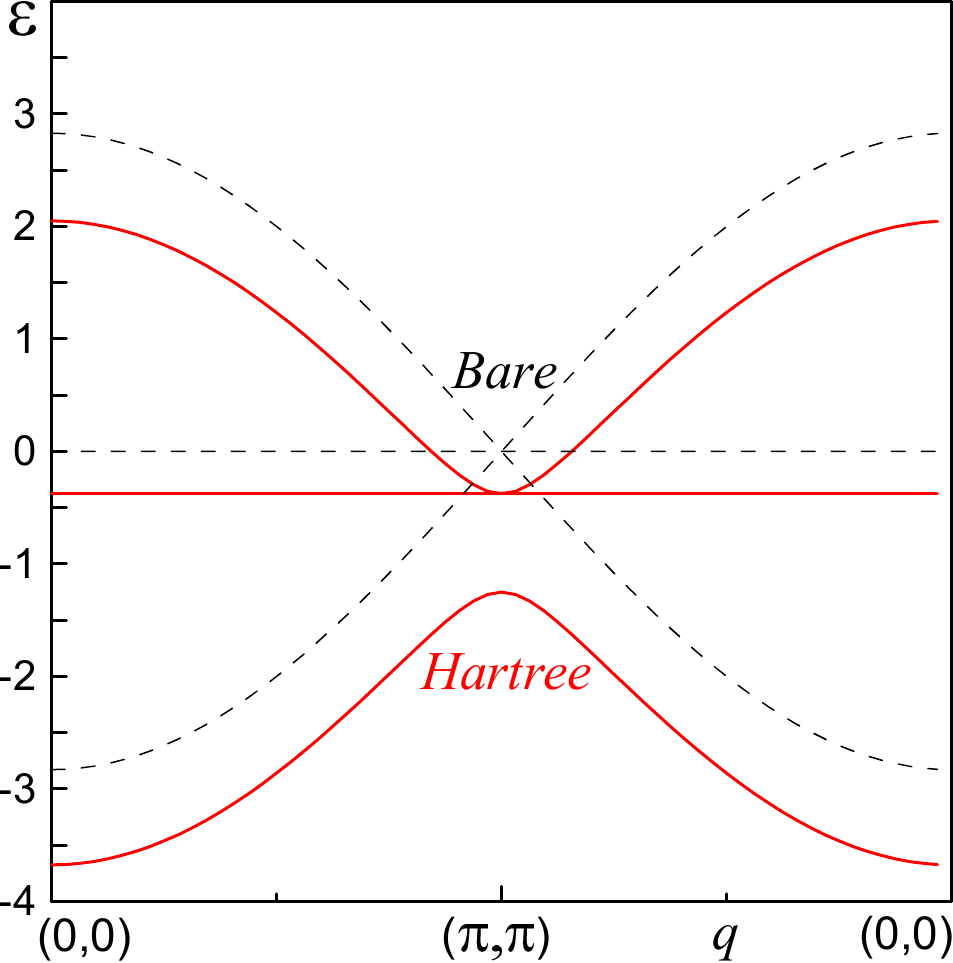}
\caption{
Left: Standard Lieb lattice with three sites $\{ A, B, C \} $ in the unit cell and hopping amplitudes $t$ between the n.n. sites. Right: Bare (black dashed line) and Hartree mean-field (solid red line) dispersion relations for $U=-2t$ at $n=2$ along the $k_x=k_y$ line in the Brillouin zone.
}
\label{fig:Lieb}
\end{figure}

\textit{Model.} 
We study an attractive Hubbard model on the standard Lieb lattice \cite{lieb_two_1989} defined by the Hamiltonian
\begin{equation}
\hat{H} =  -t\sum_{\langle\mathbf{i}a,\mathbf{j}b\rangle, \sigma}
\hat{c}_{\mathbf{i}a\sigma}^{\dagger}
\hat{c}_{\mathbf{j}b\sigma}+
U\sum_{\mathbf{i}a} \hat{n}_{\mathbf{i}a\uparrow} \hat{n}_{\mathbf{i}a\downarrow},
\label{H}
\end{equation}
where $\hat{c}_{\mathbf{i}a\sigma}^{\dagger}$ creates a fermion on a site
defined by the unit cell index $\mathbf{i}$ and the orbital index $a=\textrm{A,B,C}$ (labeling the atoms in the unit cell) with spin projection $\sigma\in\{\uparrow,\downarrow\}$, and $\hat{n}_{\mathbf{i}a\sigma}=\hat{c}^{\dagger}_{\mathbf{i}a\sigma} 
\hat{c}_{\mathbf{i}a\sigma}$. The first sum is restricted to the nearest-neighbor (n.n.) sites, see left panel in Fig.~\ref{fig:Lieb}. In this work we take the n.n. hopping amplitude, $t$, and lattice constant $a_0$, as the units of energy and distance, respectively. The bare band structure is gapless with all three bands touching at the momentum point $(\pi, \pi)$, see right panel in Fig.~\ref{fig:Lieb}. To study bound states, we consider the filling (per unit cell) $n=2$, corresponding to the lower band being fully occupied.

\textit{Results.} 
We start by noticing that when the sites inside the unit cell have different energies, e.g. $E_1$ on sublattice A is smaller than $E_2=E_3$ on sublattices B and C, then a gap, $\Delta=|E_2-E_1|$, opens below the FB \cite{tsai_interaction_2015}. This is exactly what happens in the standard Lieb lattice due to Hartree mean-field (HMF) shifts, $E_a = U n_a/2 $, when the lower band is fully occupied (here $n_a=2\langle \hat{n}_{\mathbf{i}a\sigma} \rangle$). The prediction of the self-consistent HMF for $U=-2t$ is $n_1 = 1.250$, $n_2 = n_3 =0.375$, and $\Delta = 0.875$, see right panel in Fig.~\ref{fig:Lieb}. Self-consistency is required for $U=-2t$ because the linear regime $\Delta \approx |U|/4 $ extends only up to $|U| < t/2$. By absorbing the Hartree shifts into the renormalized bare Hamiltonian,
\begin{equation}
\hat{H}_{HMF} =
-t\sum_{\langle\mathbf{i}a,\mathbf{j}b\rangle, \sigma}
\hat{c}_{\mathbf{i}a\sigma}^{\dagger}
\hat{c}_{\mathbf{j}b\sigma} +
\sum_{\mathbf{i}a,\sigma} E_{a} \hat{n}_{\mathbf{i}a\sigma},
\label{HMF}
\end{equation}
we arrive at a meaningful formulation of the bound-state problem under the assumption that the beyond-MF terms preserve the gap and the binding energy satisfies the condition $E_b > -2 \Delta$, or $\delta = E_b + 2 \Delta >0 $. In the absence of the $\Delta \gg |U|$ condition, this is not guaranteed and the problem remains non-perturbative, i.e. many-body particle-hole excitations may play an important role.

\begin{figure}[t]
\centering
\includegraphics[width=0.9\linewidth] {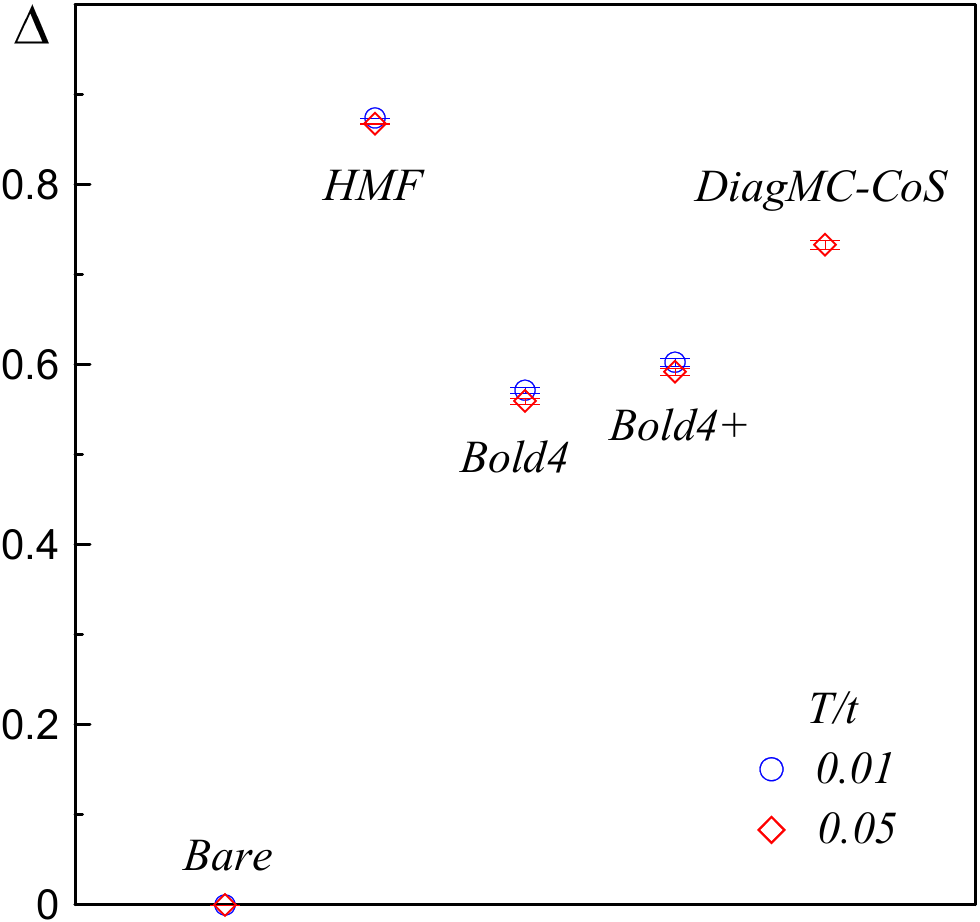}
\caption{Energy gap at momentum $(\pi, \pi)$ between the fully occupied lowest band and the empty (nearly) flat band for $U/t = -2$, shown as a function of the method used.
}
\label{Gap-app}
\end{figure}

To verify the validity of the first assumption, we employed two ``beyond-MF" diagrammatic schemes and the systematic high-order diagrammatic approach (DiagMC) with a controlled estimate of the error bars. The Bold4 scheme \cite{simkovic_magnetic_2017} is based on four self-consistent Dyson equations for the Green's function, $G$, and three two-particle propagators. The key approximation is that all four self-energies are based on the lowest-order one-loop diagrams (see Fig.~6 in Ref.~\cite{simkovic_magnetic_2017}). Bold4 is exact at the $U^2$-level and includes higher-order contributions only in the form of embedded geometrical series. Screening effects based on particle-hole excitations are already captured by this scheme. Using Bold4 results as the starting point, the leading vertex corrections are added by the Bold4+ scheme \cite{tupitsyn_numerically_2026, tupitsyn2026avoidedstonerinstabilitysingle}, which makes Bold4+ results exact at the $U^4$-level for single-particle propagators and $U^3$-level for two-particle propagators by accounting for non-local pair scattering. Finally, we performed high-order diagrammatic simulations using the combinatorial summation (CoS) algorithm \cite{Kozik2024}, which deterministically sums all $\propto (n!)^2$ diagram integrands using a computational graph and performs Monte Carlo (MC) integration over internal space-time coordinates. The value behind the series is reconstructed using Dlog Pad\'e~\cite{Pade1961} and integral approximant methods~\cite{Hunter1979IA}, following the protocol established in Ref.~\cite{PadeSK2019} and validated against other precision techniques~\cite{Schafer2021} and ultracold-atom experiment~\cite{Currie_2025validation}. In this approach, consistency among distinct extrapolations within the statistical error propagated from the series coefficients serves as an intrinsic test of error control, while any residual spread is incorporated into the reported error bar.

The single-particle gap between the fully occupied first (lowest) band and the empty second (central) band at the momentum point $(\pi, \pi)$---which we still refer to as ``flat'' despite the second band becoming weakly dispersive due to interactions (with bandwidth of about $0.03t$ at $U/t=-2$) ---was extracted from the pole of the Green's function. Figure~\ref{Gap-app} shows the results for $\Delta$ obtained from all diagrammatic methods. Although including self-consistent one-loop diagrams in the Bold4 method significantly decreases $\Delta$, incorporating higher-order vertex corrections increases the gap, which ultimately settles between the HMF and Bold4+ results.

Having established that the FB is separated from the lower band by a gap, we proceed with computing properties of the bound state formed when two fermions are introduced into the system. The simplest scheme is based on the variational two-body ansatz \cite{torma_quantum_2018, iskin_two-body_2021} that needs to be solved \textit{after} the Hamiltonian is renormalized by HMF shifts:
\begin{equation}\label{eqn:variational_ansatz}
\ket{\Psi}=\sum_{\mathbf{k}nm}A_{\mathbf{k}nm}\hat{c}^{\dagger}_{\mathbf{k},n,\uparrow}\hat{c}^{\dagger}_{\mathbf{q-k},m,\downarrow}\ket{0},
\end{equation}
where the sum over $\mathbf{k}$ runs over the first Brillouin zone, $\mathbf{q}$ is the total momentum of the pair, and $n$ is the band index. The ``vacuum" state $\ket{0}$ corresponds to the filled lower band, and thus the sum over $n$ and $m$ in Eq.~(\ref{eqn:variational_ansatz}) is restricted to the flat and upper bands. The variational ansatz (\ref{eqn:variational_ansatz}) prevents one from considering particle-hole excitations when electrons from the lower band are promoted to empty bands. This formally corresponds to solving the two-body problem with an interaction Hamiltonian acting exclusively in the subspace of empty HMF bands.

\begin{figure}[t]
\centering
\includegraphics[width=0.9\linewidth]{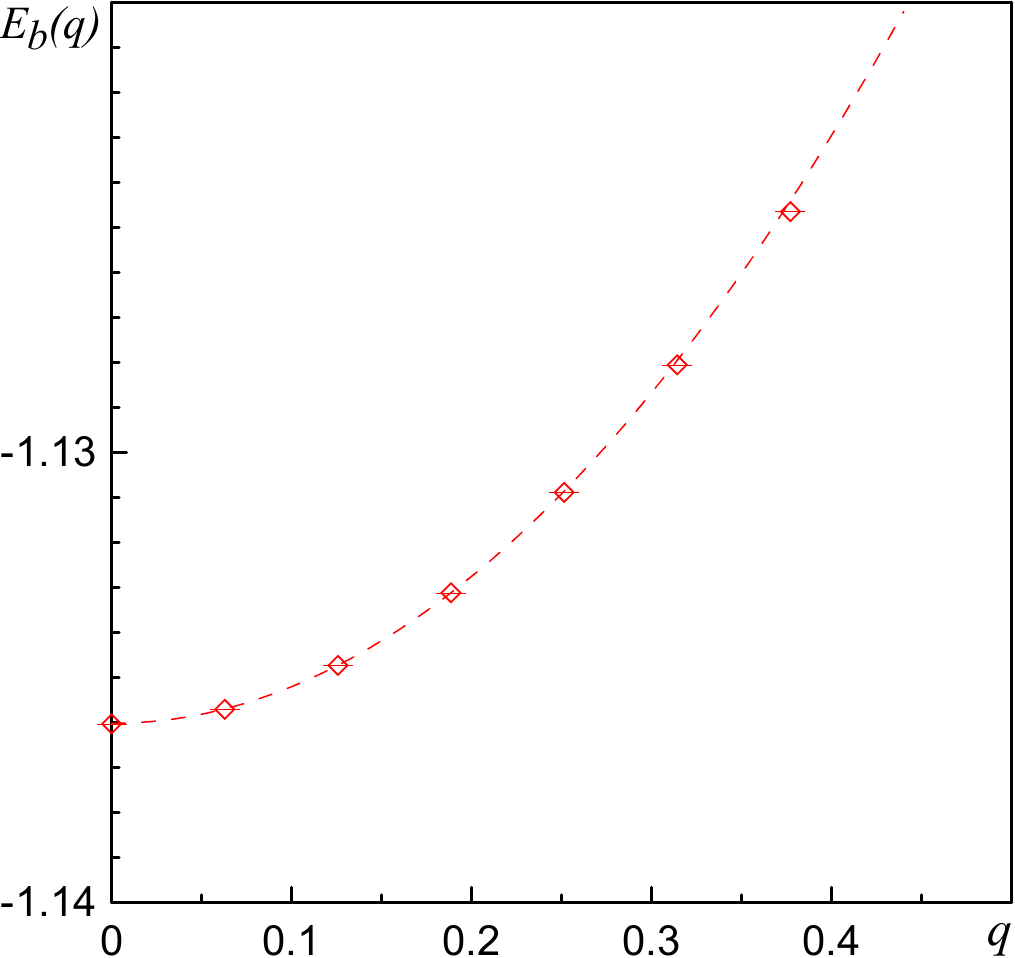}
\caption{Bound state energy obtained within the variational anzats
(\ref{eqn:variational_ansatz}) on top of the HMF-renormalized band structure
for $U=-2t$ (the numerical solution was obtained for the lattice with
$100 \times 100$ unit cells and periodic boundary conditions).
The dashed line is a parabolic fit
$E_b(q)=E_b(0)+q^2/2m_*$ with $1/m_*=0.164 \pm 0.002$.
}
\label{fig:mass}
\end{figure}

Following Ref.~\cite{iskin_two-body_2021}, the eigenvalue/eigenfunction problem for the bound state in the restricted Hilbert space can be reduced to solving
\begin{equation}\label{eqn:alpha_equation}
\alpha_a = U 
\sum_b\sum_{\mathbf{k}nm}\frac{f_{nm}^a(\mathbf{k})f_{nm}^b(\mathbf{k})^*}{E-\epsilon_n(\mathbf{k})-\epsilon_m(\mathbf{q-k})}\alpha_b\,,
\end{equation}
for the lowest value of $E=E_b(q)$. Here $f_{nm}^a(\mathbf{k})=u_n^a(\mathbf{k})u_m^a(\mathbf{q-k})$, $u_n^a(\mathbf{k})$ is the Bloch function with momentum $\mathbf{k}$ of band $n$ on sublattice $a$, $\epsilon_n(\mathbf{k})$ is the single-particle dispersion obtained by the diagonalization of (\ref{HMF}), and $\alpha_a=\sum_{\mathbf{k}nm}f_{nm}^a (\mathbf{k})A_{\mathbf{k}nm}$. The result of this calculation is presented in Fig.~\ref{fig:mass}. It shows that the bound state is located withing the gap, i.e.
$\delta > 0$, and has an effective mass $m_* \approx 6.1$ (in units of $1/ta_0^2$).

To investigate bound-state properties within the standard grand-canonical diagrammatic framework, we work at sufficiently low temperatures and set the chemical potential within the interval $(E_b/2, -\Delta)$---with energies referenced to the bottom of the central band---such that the conditions $T/(\Delta + \mu) \ll 1$ and $T/(E_b - 2\mu) \ll 1$ are met. The first condition is necessary to keep the lowest band fully occupied. The second, more restrictive, condition is required for keeping the density of bound pairs exponentially low. The calculation then proceeds with computing the pair propagator
\bea
G_{pp} (\mathbf{q}, \tau ) =
\sum_{\mathbf{i},ab} e^{i\mathbf{q}\cdot \mathbf{i} }
\langle
\hat{c}_{\mathbf{i}b\downarrow} (\tau )
\hat{c}_{\mathbf{i}b\uparrow}  (\tau )
\hat{c}_{0a\uparrow}^{\dagger}(0)
\hat{c}_{0a\downarrow}^{\dagger}(0)
\rangle ,
\label{gpp}
\eea
and extracting $E_b(\mathbf{q})$ from its asymptotic decay
\begin{equation}
\ln \left[ G_{pp} (\mathbf{q}, \tau \to \infty ) \right] =
{\rm const} -(E_b(\mathbf{q})-2\mu ) \tau .
\label{logGpp}
\end{equation}

\begin{figure}[t]
\centering
\includegraphics[width=0.95\linewidth] {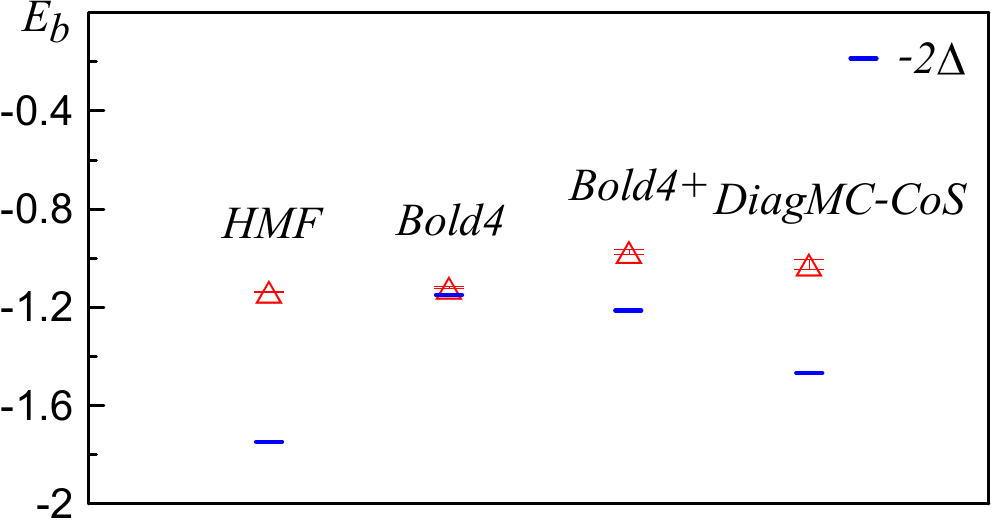}
\caption{
Binding energy predictions (red triangles) across several schemes with an increasingly comprehensive treatment of particle-hole excitations. Blue dashed lines show the corresponding $-2\Delta$ values. Results are shown for $U/t = -2$, with Bold4 and Bold4+ simulations performed at $T/t = 0.01$ and DiagMC-CoS simulations at $T/t = 0.05$.
}
\label{E0-app}
\end{figure}

Results for the binding energy, $E_b\equiv E_b(q=0)$, from Bold4, Bold4+, and DiagMC-CoS calculations are presented in Fig.~\ref{E0-app}. While, similarly to HMF, higher-order diagrammatic schemes also find that the condition $E_b > -2\Delta $, is satisfied, i.e. they predict that the IIBI state is stable, screening effects from the occupied band accounted for at the leading diagrammatic orders within the Bold4 scheme bring the system extremely close to the point of instability. In Bold4, the value of $\delta$ is only about a few percent of $|E_b|$, see the difference between the data points and dash lines in  Fig.~\ref{E0-app}. This indicates that the problem remains highly non-perturbative beyond HMF. However, when the leading (and higher-order) vertex corrections are accounted for within the Bold4+ and DiagMC-CoS schemes, the value of $\delta$ increases, and the system's stability at $n=2$ is no longer in doubt. Variations in $\delta$ are the most significant many-body effect because the value of $E_b$ itself barely changes by about $10$\% between the schemes.

\begin{figure}[t]
\centering
\includegraphics[width=0.95\linewidth] {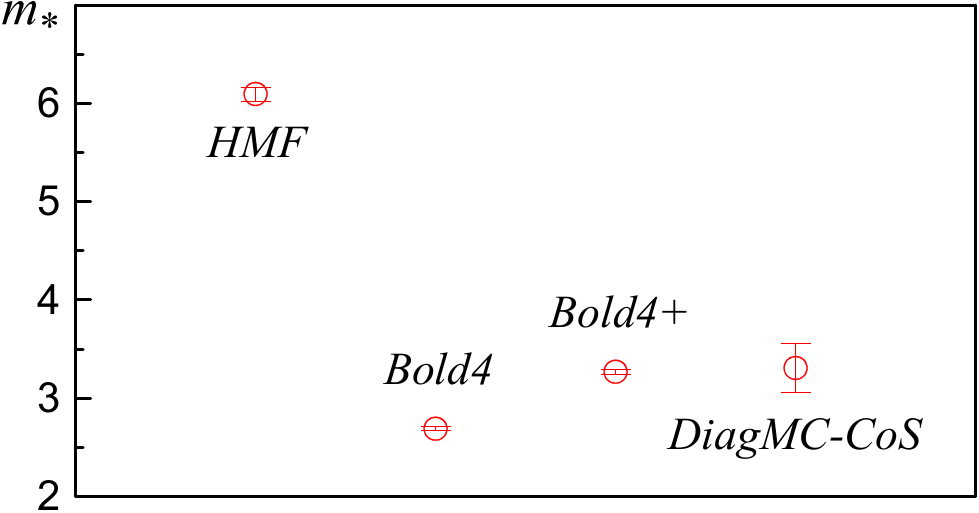}
\caption{
Effective mass predictions using the same computational schemes and parameters as in Fig.~\ref{E0-app}.
}
\label{Mass-app}
\end{figure}

Parabolic dependence of the binding energy on momentum at $q \to 0$ provides an estimate of the pair effective mass, $m_*$; in our case, the effective mass tensor is diagonal and isotropic. In Fig.~\ref{Mass-app} (see also Fig.~\ref{fig:mass}), we show the dependence of $m_*$ on the approach used to compute it. For this property, the difference between the HMF and diagrammatic
schemes is at the level of $100$\%. It appears that the most significant effect leading to a much lighter effective mass is already captured by the Bold4 scheme, which accounts for additional mechanisms of moving a pair. For example, the interaction Hamiltonian allows transitions that create holes in the occupied band and particles in the upper band, propagation of particles in the upper band, and subsequent recombination of holes with flat-band particles. After minor change in $m_*$ between the Bold4 and Bold4+ schemes, we quickly obtain a converged result as evidenced by agreement (within the error bars) between the Bold4+ and DiagMC-CoS calculations, see Fig.~\ref{Mass-app}.

\textit{Conclusions.}
We have found that the bosonic---or preformed Cooper-pair---mechanism can drive flat-band superconductivity even when the bare band structure is gapless. In the absence of a small parameter, properly accounting for correlations arising from hole excitations in the lower occupied band becomes essential, and we capture these effects using a hierarchy of advanced diagrammatic schemes. In the revealed scenario, interactions open a single-particle gap between the lower occupied band and the empty flat band (which becomes only weakly dispersive) and bind fermions doped into the flat band into highly mobile pairs.

\begin{figure}[h]
\centering
\includegraphics[width=0.9\linewidth] {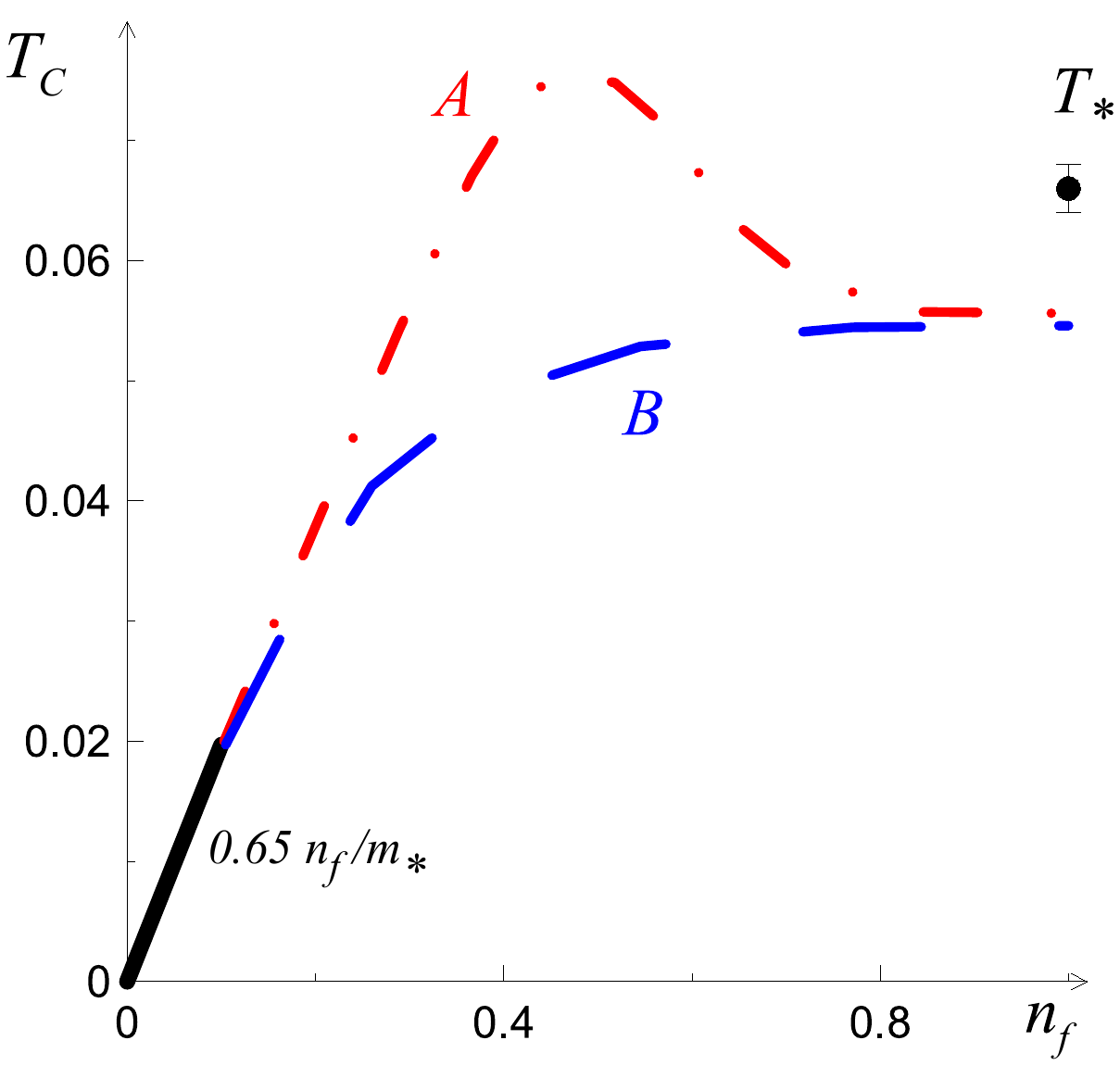}
\caption{
Possible phase diagrams for the superconducting transition temperature dependence on density of fermions doped into the flat band of the standard Lieb lattice for attractive on-site interaction $U=-2t$. Dashed and dot-dashed lines are different possible scenarios for connecting low-density dilute Bose gas mechanism to the $n_f=1$ result calculated in Ref.~\cite{tupitsyn_numerically_2026}.
}
\label{fig:schematicTc}
\end{figure}

Figure~\ref{fig:schematicTc} illustrates two candidate scenarios for the evolution of the critical temperature $T_c$ as a function of the flat-band fermion density $n_f$. In the dilute limit ($n_f \ll 1$), $T_c(n_f)$ is dictated by a Berezinskii-Kosterlitz-Thouless (BKT) transition in a dilute Bose gas of pairs and follows the black line whose slope is set by the pair mass $m_* \approx 3.3 / (t a_0^2)$ obtained here. At half-filling ($n_f = 1$), where the preformed-pair description no longer applies, $T_c$ remains bounded by $T_*$, the crossover scale for the onset of long-range superconducting response established in Ref.~\cite{tupitsyn_numerically_2026}. Given these two boundary behaviors, $T_c(n_f)$ could either develop a non-monotonic peak (case A) or quickly saturate (case B). While resolving this behavior warrants further study, in either scenario the high critical temperatures identified in Ref.~\cite{tupitsyn_numerically_2026} should remain robust over a broad density range near half-filling.

\begin{acknowledgments}
\textit{Acknowledgments.} VB, IST, BVS, and NVP acknowledge support from the Simons Foundation grant SFI-MPS-NFS-00006741-07 for the Simons Collaboration on New Frontiers in Superconductivity. BC, EK, BVS and NVP acknowledge support from EPSRC through Grant No. EP/X01245X/1. The DiagMC calculations were performed using King's Computational Research, Engineering and Technology Environment (CREATE). This work used the ARCHER2 UK National Supercomputing Service (https://www.archer2.ac.uk) \cite{Archer}.
\end{acknowledgments}

\bibliography{refs.bib}

\end{document}